\documentclass[a4paper]{article}

\usepackage{ISCSLP2022}

\title{CUEMPATHY: A Counseling Speech Dataset for Psychotherapy Research}
\name{Dehua Tao$^1$, Harold Chui$^2$, Sarah Luk$^2$, Tan Lee$^1$}
\address{
  $^1$ Department of Electronic Engineering \quad 
  $^2$ Department of Educational Psychology\\The Chinese University of Hong Kong}
\email{dhtao@link.cuhk.edu.hk, tanlee@ee.cuhk.edu.hk, \{haroldchui, sarah\_luk\}@cuhk.edu.hk}

\begin{document}

\maketitle
\begin{abstract} % limited to 200 words
Psychotherapy or counseling is typically conducted through spoken conversation between a therapist and a client. Analyzing the speech characteristics of psychotherapeutic interactions can help understand the factors associated with effective psychotherapy. This paper introduces CUEMPATHY, a large-scale speech dataset collected from actual counseling sessions. The dataset consists of 156 counseling sessions involving 39 therapist-client dyads. The process of speech data collection, subjective ratings (one observer and two client ratings), and transcription are described. An automatic speech and text processing system is developed to locate the time stamps of speaker turns in each session. Examining the relationships among the three subjective ratings suggests that observer and client ratings have no significant correlation, while the client-rated measures are significantly correlated. The intensity similarity between the therapist and the client, measured by the averaged absolute difference of speaker-turn-level intensities, is associated with the psychotherapy outcomes. Recent studies on the acoustic and linguistic characteristics of the CUEMPATHY are introduced.
\end{abstract}

\section{Introduction}
\label{sec:intro}

Psychotherapy or counseling is an activity of conversational speaking involving a client and a therapist/counselor. ``Through the verbal transactions of the therapy" \cite{wampold2012humanism}, clients are encouraged to express their thoughts and feelings, think more deeply about issues at hand, and make changes in life. The aims of psychotherapy are to lower clients' psychological distress, enhance their well-being, and reduce maladaptive behaviors that hinder their relationship and work functioning \cite{hill2020helping}. Better understanding about the factors associated with effective psychotherapy can be achieved by examining the speech characteristics of psychotherapeutic interactions in conjunction with traditional assessments of psychotherapy process and outcome. Previous studies investigated how language expression and style are correlated with psychotherapy outcomes \cite{xiao2016behavioral,lord2015more,gibson2015predicting,chakravarthula2015assessing,xiao2012analyzing}. Vocal cues related to the counseling quality were studied extensively in \cite{de2019investigation,xiao2015analyzing,xiao2014modeling,imel2014association,xiao2013modeling}. The findings of these studies are useful to the prediction of clinical outcomes and can guide what and how therapists should speak and express during counseling to maximize therapeutic effect.

One important feature of counseling and psychotherapy is confidentiality. Protecting clients' privacy does not only aim to meet ethical and professional standards \cite{aca2014,ethical2017}, but also help clients build up trust so that they can work on their issues freely and productively. To uphold the highest ethical practice, the standard of confidentiality extends beyond the counseling room to research activities. In the context of applying the latest data analytics technology, special attention must be paid when psychotherapy recordings need to be transmitted over the network for processing by any third-party software, e.g., cloud-based speech transcription, given the uncertainty around data storage and use. Psychotherapy researchers therefore choose to collect their own data and use these data to build tailor-made systems for data analysis.

This paper describes a recent study on constructing a large-scale database of counseling speech. The data are collected from actual counseling sessions at a counseling clinic of the Department of Educational Psychology in our university. In each session, a counseling trainee (master student) is arranged to see a help-seeking client under supervision. The speech dataset, named CUEMPATHY, contains hundreds of hours of audio recordings and text transcriptions of spoken conversation between therapists and clients. These data are intended to support the development of speech and language technologies that can be applied to analyze psychotherapy interactions efficiently and provide data evidences for counselor training.

Section \ref{sec:datacol} gives the background and describes the process of data collection. Section \ref{sec:ratingtrans} details how manual transcription and empathy rating are done. Section \ref{dataproc} presents what has been done on the processing of raw recordings. Section \ref{sec:preinves} and \ref{sec:work} describe our preliminary investigation and on-going work on counseling speech analysis with CUEMPATHY. Conclusion is given in Section \ref{sec:conc}.

\section{Data Collection}
\label{sec:datacol}

\subsection{Background}
\label{subsec:back}

% (This study was conducted in a departmental clinic where counseling practicum students received training at a university in Hong Kong, China. Trainees enrolled in the 20-week-long practicum as part of their master’s program. Practicum clients included adults and children from the community who consulted with the clinic for reduced-fee counseling over a range of concerns related to stress, relationship, career, emotions, self-esteem, and personal growth. Trainees do not follow specific treatment protocol but discuss with supervisors the appropriate counseling goals and strategies. For this study, only adult clients in individual therapy participated, and data were collected over four cohorts of practicum students between November 2017 and October 2019. Before counseling, potential clients participated in a telephone screening and were referred to other health professionals in the community if they were deemed inappropriate to be seen by novice counselors. Reasons for referral included ongoing suicidal ideation, psychosis, active substance use, or any other concerns requiring more intensive intervention. Trainees received group and individual supervision as part of training.)

In the psychotherapy practicum, counseling trainees are required to take part in a training program of 20 weeks. The clients are adults from the general community who come to seek reduced-fee psychotherapy services. The clients' concerns are related to emotion, stress, career, relationship, personal growth, and self-esteem. All trainee therapists and clients speak Hong Kong Cantonese with occasional English code-mixing. An informed consent was obtained from each participating client or trainee therapist. The procedures followed the standard research ethics of the American Psychological Association and were approved by the university's Institutional Review Board.

Before formal counseling, potential clients were asked to go through a telephone screening. If they were deemed unsuitable for the service, they would be referred to other health professionals in the community. Reasons for referral included ongoing suicidal ideation, psychosis, active substance use, or any other concerns requiring intensive intervention. For the enrolled clients, trainee therapists discussed with their supervisors the counseling goals and strategies rather than following a specific treatment protocol. Counseling sessions were conducted weekly during the 20-week practicum, but not every client started in the first week, and not every client ended in the last week. As part of training, trainee therapists received individual and group supervision.

\begin{table}[htb]
\centering
\caption{Gender and age of therapists and clients in the CUEMPATHY}
\label{tab:spkinfo}
\resizebox{1.0\linewidth}{!}{
\begin{tabular}{|c|c|c|c|}
\hline
          & Female & Male & Age (range; mean \textpm std) \\ \hline
Therapist & 31     & 8    & 24 to 58; 34.26 \textpm 7.96    \\ \hline
Client    & 30     & 9    & 18 to 57; 34.59 \textpm 10.82   \\ \hline
\end{tabular}
}
\end{table}

A total of 428 counseling sessions were recorded with 4 cohorts of practicum trainees (a total of 40 trainee therapists) during the period of November 2017 to October 2019. Each of the 40 therapists was paired with a designated client. Thus there were totally 40 unique therapist-client dyads. The number of counseling sessions conducted by these dyads varied from 6 to 16, with the mean of $10.7\pm2.83$. 

The CUEMPATHY dataset is made up of 156 selected sessions. They cover 39 therapist-client dyads (the excluded dyad has no audio recording) and 4 sessions per dyad. Table \ref{tab:spkinfo} gives the gender and age information of therapists and clients.

\subsection{Recording session setup}
\label{subsec:recset}

All counseling sessions were conducted in a sound-attenuated room. Only the therapist and the client were present in the room. Each session was about 50 minutes long. All sessions were video-taped by a camera mounted on the roof and audio-recorded by a digital recorder (TASCAM SD-20M). The therapist and client each wore a lavalier microphone clipped between the collar and chest. The microphones were connected to the recorder. Audio recording was done with sampling rate of 48 kHz and two channels (stereo). Before a session commenced, a research assistant helped to install the microphones for both speakers and tested the recording equipment. In a typical session, the therapist and the client took turn to speak. A speaker turn refers to the time period in which only one person speaks.

\section{Empathy Rating and Speech Transcription}
\label{sec:ratingtrans}

\subsection{Subjective measurement of empathy}
\label{subsec:meas}

Therapist empathy has long been hypothesized to be a primary indicator of counseling outcomes \cite{miller2009toward,elliott2011empathy,moyers2013low,elliott2018therapist}. Empathy is described as ``the therapist's sensitive ability and willingness to understand the client's thoughts, feelings, and struggles from the client's point of view" \cite{rogers1995way}. Two instruments are used to measure the therapist empathy in this research. They include Therapist Empathy Scale (TES) \cite{decker2014development} rated by third-party observers, and Barrett-Lennard Relationship Inventory (BLRI) \cite{barrett2015relationship} rated by the clients. Another client-rated measure, Session Evaluation Scale (SES) \cite{hill2002development}, is also adopted to assess clients' perception of session quality. The session quality measures the effectiveness of counseling session from client's perspective.

\subsubsection{Therapist Empathy Scale (TES)}
\label{subsubsec:tes}

TES is an observer-rated measure of therapist empathy. It comprises nine items that cover the affective, cognitive, attitudinal, and attunement aspects of empathy. The observer gives a score on each item after watching the video recording of each session. The scores follow a 7-point scale: $1 = \textit{not at all}$ to $7 = \textit{extremely}$.  An example item is ``Concern: A therapist conveys concern by showing a regard for and interest in the client. The therapist seems engaged and involved with the client and attentive to what the client has said. The therapist's voice has a soft resonance that supports and enhances the client's concerned expressions.'' The total score (ranging from 9 to 63) is used in this research, with a higher value indicating higher therapist empathy.

% (The TES is an observer rating scale to assess affective, cognitive, attitudinal, and attunement aspects of therapist empathy in a therapy session. This scale consists of nine items: therapist concern, expressiveness, resonate or capture client feelings, warmth, attuned to client's inner world, understanding cognitive framework, understanding feelings/inner experience, acceptance of feelings/inner experiences, and responsiveness. Each item is rated on a seven-point scale from $1 = \textit{not at all}$ to $7 = \textit{extremely}$. A score is given to each item after observers finish watching a videotaped counseling session. An example item is ``Concern: A therapist conveys concern by showing a regard for and interest in the client. The therapist seems engaged and involved with the client and attentive to what the client has said. The therapist's voice has a soft resonance that supports and enhances the client's concerned expressions." The total score (range from 9 to 63) is used in this research, with a larger value indicating higher observer-rated therapist empathy. The TES has excellent psychometric properties in the scale development sample \cite{decker2014development}, and its internal consistency for the current sample is high ($\alpha = 0.96$).)

\subsubsection{Barrett-Lennard Relationship Inventory (BLRI)}
\label{subsubsec:blri}

Being a client-rated measure of therapist empathy, the BLRI is made up of sixteen items. Each item is rated on a 6-point Likert-type scale from $-3 = \textit{strongly disagree}$ to $+3 = \textit{strongly agree}$. Zero value is not used in the scoring. An example item is: ``My counselor tries to see things through my eyes.'' The overall BLRI score ranges from $-48$ to $+48$, with a higher value indicating higher therapist empathy as perceived by clients.
% The Cronbach’s $\alpha$ for the present study is 0.82.

\subsubsection{Session Evaluation Scale (SES)}
\label{subsubsec:ses}

In our research, SES consists of five items, including four items from the original SES \cite{hill2002development} rated on a 5-point Likert-type scale from $1 = \textit{strongly disagree}$ to $5 = \textit{strongly agree}$, and one added item for assessing session effectiveness and increasing scale variance as suggested by Lent \textit{et al.} \cite{lent2006client}. An example item is: ``I am glad I attended this session.'' The value of SES ranges from 5 to 25. A higher value indicates better session quality as perceived by clients.
% The Cronbach’s $\alpha$ for the present study is 0.68.

% (The original SES included four items rated on a 5-point Likert-type scale from $1 = \textit{strongly disagree}$ to $5 = \textit{strongly agree}$. A fifth item was added to assess overall session effectiveness and increase scale variance, as suggested by Lent \textit{et al.} \cite{lent2006client}. The SES scores can range from 5 to 25. A larger value indicates better session quality perceived by the clients.)

\subsection{Transcription and empathy rating of CUEMPATHY}
\label{subsec:trans}

Speech transcription was done on a speaker-turn basis, with the speaker identity marked, i.e., therapist or client, and the speech content transcribed in the form of traditional Chinese characters with punctuation marks. The transcription was done by trained undergraduate research assistants at the counseling clinic. Identifiable information, such as names and locations, were removed from the transcriptions before analysis. Only research personnel involved in this study are allowed to access the audio data and text transcriptions.

Clients were asked to complete both BLRI and SES after each session. For observer-rated TES, 8 raters with at least master’s level training in counseling were recruited. They were trained first on performing TES rating with 8 video-taped counseling sessions that had been collected with the same setting. The intra-class coefficient (ICC) for the 8 raters on the 8 training sessions was $0.79$. According to the Cicchetti’s guidelines \cite{cicchetti1994guidelines}\footnote{The Cicchetti’s guidelines on inter-rater reliability: ICCs $< 0.40$ mean poor, $0.40 - 0.59$ mean fair, $0.60 - 0.74$ mean good, and $> 0.75$ mean excellent reliability.}, the raters were considered to achieve excellent inter-rater reliability. The raters then proceed to rate the sessions in CUEMPATHY. As a reliability check, about $40\%$ (62 sessions) of the sessions were rated by two raters. The ICC based on a mean-rating (k = 2), consistency, two-way random effects model was $0.90$, indicating excellent inter-rater reliability beyond the training phase.

% (In the present study, eight raters were first trained on the TES using eight videotaped counseling sessions unrelated to the study sample but were collected from the same setting. Based on these sessions, the intraclass coefficient (ICC) for the eight raters was .79. Using Cicchetti (1994)’s guidelines on interrater reliability, where ICCs $< .40$ are poor, $.40 - .59$ are fair, $.60 - .74$ are good, and $> .75$ are excellent, the raters are considered to have excellent interrater reliability. The raters then proceeded to rate counseling sessions from the study. As a reliability check, about 40\% (61 sessions) of the videotapes were rated by two raters. The ICC based on a mean-rating (k = 2), consistency, two-way random effects model was .90, indicating excellent interrater reliability beyond the training phase.)

% (Raters all have at least master’s level training in counseling, and most of them are female. Out of 160 sessions, 94 were rated by single raters, and 66 were rated by two or more raters. We added a third rater when the difference in scores between the raters were large, and we took the average of the closer of the two raters. For 72 of the sessions, the same rater rated four sessions from the same dyad.)

\section{Processing of Speech Recordings}
\label{dataproc}

\begin{figure*}[htb]
\centering
\centerline{\includegraphics[width=\linewidth]{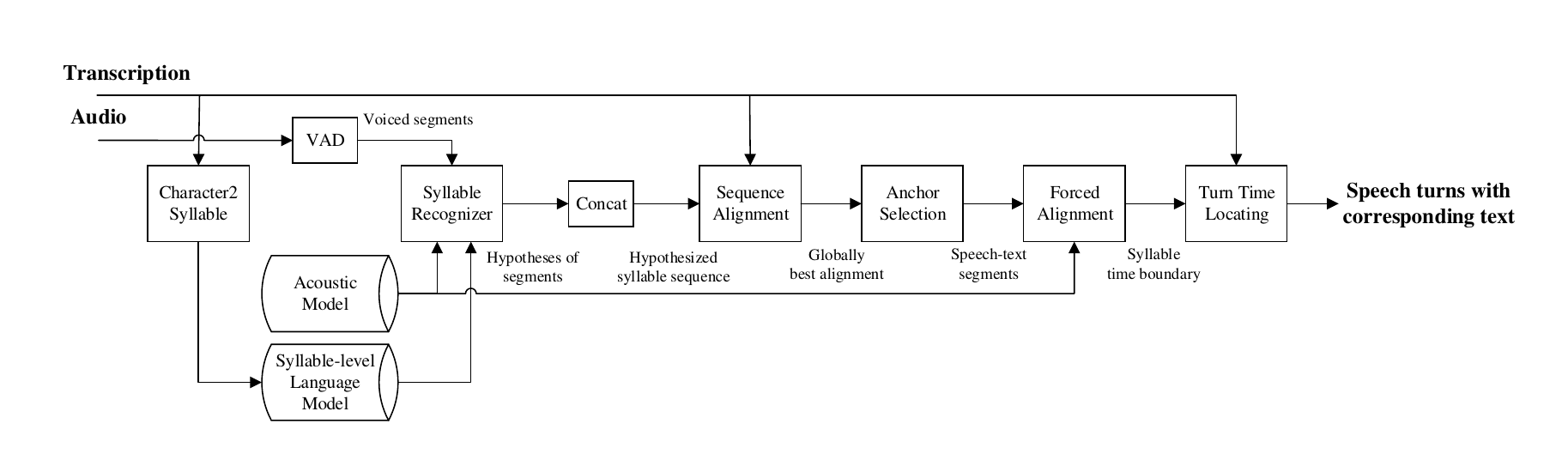}}
\vspace{-1.0em}
\caption{Block diagram of the turn-level speech-text alignment system.}
\label{fig:aligner}
% \vspace{-1.0em}
\end{figure*}

\begin{table*}[htb]
\caption{Summary of the speech data in CUEMPATHY}
\label{tab:data}
\centering
\resizebox{1.0\textwidth}{!}{
\begin{tabular}{|l|c|c|c|c|c|}
\hline
          & \begin{tabular}[c]{@{}c@{}}Average speech time \\ per session (min)\end{tabular} & \begin{tabular}[c]{@{}c@{}}Average no. of characters \\ per session\end{tabular} & \begin{tabular}[c]{@{}c@{}}Average no. of speaker turns\\ per session\end{tabular} & \begin{tabular}[c]{@{}c@{}}Average duration \\ of turns (sec)\end{tabular} & \begin{tabular}[c]{@{}c@{}}Average no. of characters \\ per turn\end{tabular} \\ \hline
Therapist & 15.04                                                                             & 3750                                                                             & 158                                                                                & 6.38                                                                       & 26                                                                            \\ \hline
Client    & 33.42                                                                             & 7800                                                                             & 158                                                                                & 16.40                                                                      & 64                                                                            \\ \hline
\end{tabular}
}
\end{table*}

In order to facilitate focused analysis of therapist and client speech, the time stamp information of each speaker turn in the audio recording needs to be known. Since the speaker-turn-based transcriptions are available, this problem can be formulated as (1) aligning long audio to corresponding text transcription \cite{moreno1998recursive}; (2) finding the beginning and ending time of all speaker turns.

Inspired by the work of \cite{moreno1998recursive}, a speech-text alignment system is developed to produce turn-level time alignment of the recorded counseling speech. Figure \ref{fig:aligner} shows the system's block diagram. In the following sub-sections, we will explain the process of obtaining turn-level alignment of speech data and text transcription for a 50-minute recording.

\subsection{Data pre-processing}
\label{subsec:preproc}

The original stereo audio is first converted into mixed-mono audio and down-sampled to 16 kHz. The long audio is split into short voiced segments using a voice activity detection (VAD) tool, namely WebRTC VAD\footnote{https://github.com/wiseman/py-webrtcvad}. The \textit{Character2Syllable} module converts the given transcription from Chinese characters to Cantonese syllables in the form of Jyutping symbols \cite{lee2002spoken}. Code-mixing English words are also represented by Jyutping symbols that resemble the English pronunciations as closely as possible \cite{chan2009automatic}.

\subsection{Syllable recognition}
\label{subsec:sylrec}

A Cantonese syllable recognition module is built with the Kaldi toolkit \cite{povey2011kaldi}. A tri-gram language model is trained using the syllable transcription of the session. Such a session-specific language model is expected to effectively restrict the decoding search space by pruning irrelevant hypotheses. An initial acoustic model trained with other Cantonese databases is used. Each voiced segment detected by the VAD module is passed through the recognizer to produce a hypothesized sequence of syllables (in terms of Jyutping symbols). All segment-level hypotheses are concatenated chronologically to form a sequence of syllables for the whole recording.

\subsection{Sequence alignment and anchor selection}
\label{subsec:sylalign}

The hypothesized syllable sequence obtained for the entire session is aligned with the manual transcription. The globally best alignment is determined by applying dynamic programming. The aligned sub-sequences of syllables are chosen as anchors. Each anchor contains a sequence of $N$ or more consecutive syllables ($N$ is set to $10$ in our work) and is assumed to be correctly aligned. The anchors are used to partition the transcription and the audio recording into short segments. After obtaining speech segments and the corresponding text for all sessions, a new acoustic model is trained. The new model would replace the existing one for syllable recognition. The above steps are repeated with the in-domain acoustic model and speech-text segments updated iteratively.

\subsection{Forced alignment and turn time locating}
\label{subsec:forcedalign}

Based on segment-level alignment of speech and text, syllable-level time stamps, i.e., the beginning and ending time of each syllable, are obtained by the method of forced alignment. The time stamp of each speaker turn in the recording can be determined by locating the first and last syllables in the turn. So far, the speech data and text of each speaker turn in the recording are obtained. 

After processing the recordings of all counseling sessions with the proposed system, the reliability of turn-level alignment of the speech data and text transcription is manually checked. We randomly select one session from each of 39 therapist-client dyads, i.e., a total of 39 sessions. Based on the number of speaker turns, a session is divided into three sections. In each section, at least 20 consecutive speaker turns are selected to check if the turn's speech data and corresponding text are consistent. The result indicates that the reliability of turn-level speech-text alignment is sufficient for further analysis. Table \ref{tab:data} summarizes speech data in the CUEMPATHY.

\section{Preliminary Investigation on CUEMPATHY}
\label{sec:preinves}

\subsection{Relations among the three subjective ratings}
\label{subsec:relmea}

It was found that some items of the BLRI and SES in some sessions were missing. If the amount of missing items for a measure was less than $20\%$, each missing item's value was replaced by the average of the available observed values to preserve session-level information \cite{schafer2002missing}. Five sessions, each missing one item of the SES, were handled this way. All items of the BLRI and SES for one session were missing, so the session was discarded, resulting in a total of 155 sessions for the analysis of relationships among the three measures. Table \ref{tab:measures} shows the mean, standard deviation, and range of the three measures for the 155 sessions.

% (When the amount of missing data for a measure was limited, i.e., less than 20\% of a measure was incomplete, each missing value was replaced by the average of the available observed values to preserve session-level information (Schafer \& Graham, 2002). Of 624 items of BLRI and 195 items of SES, one (.16\%) measure of BLRI and five (2.56\%) measures of SES administered in 156 sessions were missing and handled this way respectively. However, all items of the BLRI and SES were missing in one session, so that session was discarded for analysis in the current study, resulting in a total of 155 sessions for actual analysis.)

\begin{table}[htb]
\centering
\caption{Session-level Mean, Standard Deviation (SD) and range of three measures}
\label{tab:measures}
\resizebox{1.0\linewidth}{!}{
\begin{tabular}{|c|c|c|c|c|}
\hline
Measure & Mean  & SD    & Max  & Min  \\ \hline
TES     & 38.77 & 7.89  & 56.5 & 18   \\ \hline
BLRI    & 16.04 & 10.77 & 46   & -17  \\ \hline
SES     & 19.12 & 2.74  & 25   & 8.75 \\ \hline
\end{tabular}
}
\end{table}

The Pearson's correlations ($\rho$) among session-level TES, BLRI, and SES scores are computed. Results show that there is no significant correlation between observer-rated TES and client-rated BLRI ($\rho = -0.07$) or SES ($\rho = -0.07$), while the BLRI and SES are strongly positively correlated ($\rho = 0.72$). The discrepancy between observer and client ratings indicates that different raters pay attention to different aspects of the counseling session in their rating decisions. For example, previous studies suggested that client ratings of therapist empathy may be based on relationship factors other than therapists' particular skills in empathic reflection \cite{elliott2011empathy, gurman1977patient}.
% Confirm the references

\subsection{Prosodic similarity between therapist and client in counseling conversations}
\label{subsec:prosim}

It is often observed that people interacting in a conversation exhibit similar speech patterns in terms of prosody, speech sounds, and lexicon \cite{de2011measuring,edlund2009pause,pardo2006phonetic}. This phenomenon has been described in a range of terms such as synchrony, entrainment, alignment, and accommodation. The similarity of interpersonal interaction has been found to be associated with empathy \cite{arizmendi2011linking,decety2004functional,preston2002empathy}. In particular, the relationship between similarity in speech prosody and therapist empathy has been extensively studied \cite{de2019investigation, xiao2015analyzing, imel2014association, xiao2013modeling}. In this study, we investigate the relations between therapist-client similarity and observer-rated empathy ratings (i.e., TES) based on two prosodic parameters, i.e., pitch and intensity.

The method proposed in \cite{xiao2015analyzing} is adopted in our analysis. The averaged absolute difference of the turn-level parameter between the therapist and the client is computed to measure the degree of entrainment. Then the correlation between session-wise difference and empathy ratings is calculated. Consider a counseling session that contains $N$ speaker turns. Since our focus is to observe the therapist's response to the client's behavior, the client's turn is followed by the therapist's turn for computing difference. Let $x(i)$ and $x(i+1)$ be parameters of speaker turn $i$ and $i + 1$ that belong to the client and the therapist, respectively. In order to remove the bias of the individual baseline, $x(i)$ for each speaker is a zero-mean parameter by subtracting the mean of raw turn-wise parameters belonging to that speaker in the session. The averaged absolute difference $D_x$ of the session is defined as in Eq. (\ref{eq}).

\begin{equation}
\begin{aligned}
  D_x = \frac{1}{N/2}\sum_{i=1}^{N/2} |x(2i) - x(2i - 1)|
  \label{eq}
\end{aligned}
\end{equation}

The Pearson's correlation $r$ between the session-wise $D_x$ and empathy ratings is computed for each parameter. A significant correlation is found in the intensity with $r = -0.17 $ and $p< 0.05$, which indicates that the higher entrainment of intensity between the client and the therapist is associated with higher empathy. The correlation coefficient of the pitch is also in a negative value, although not significant.

\section{Recent Works on CUEMPATHY}
\label{sec:work}

Tao \textit{et al.} proposed to characterize the therapist's speaking style related to empathy with the prosodic cues from therapist speech \cite{tao22_interspeech}. They found that ``the empathy level tends to be low if therapists speak long utterances slowly or speak short utterances quickly, while high if therapists talk to clients with a steady tone and volume". The findings can guide how therapists should speak during counseling. In addition, another work \cite{tao22b_interspeech} of Tao \textit{et al.} introduced a hierarchical attention network with two-level attention mechanisms to evaluate therapist empathy from the acoustic features of conversational speech in counseling sessions. This study suggested that consecutive turns (2 to 6) may jointly contribute to determining the level of therapist empathy, and observer rating of empathy tended to take the whole counseling session into consideration.

Meanwhile, Lee \textit{et al.} examined the relationships between durational patterning at discourse boundaries and client-rated therapist empathy in counseling \cite{lee22f_interspeech}. The results suggested that ``low-order temporal cues of prosodic phrasings, such as the duration of utterance-final syllables, silent pause, and speech rate, contribute to clients’ higher-order perceptual process of therapist empathy". In another work \cite{lee2022discrepancy}, Lee \textit{et al.} argued that the discrepancy in observer and client ratings of therapist empathy might be explained by therapists’ use of particles. The study showed that particle usage significantly affected observer-rated TES (the higher the particle usage, the lower the predicted TES) but not client-rated BLRI or SES.

\section{Conclusions}
\label{sec:conc}
This paper presents a speech dataset CUEMPATHY of manually transcribed and subjectively rated counseling sessions. Observer-rated TES and client-rated BLRI and TES measure the therapist empathy and the session quality. The speech data and text transcriptions of speaker turns in each session are obtained using an automatic speech-text alignment system. Preliminary investigation of the dataset suggests that (1) observer (TES) and client ratings (BLRI or TES) are not correlated, while client-rated measures are significantly correlated; (2) the degree of intensity entrainment between the therapist and the client, measured by the averaged absolute difference of speaker-turn-level intensities, is associated with empathy level. Recent findings on the CUEMPATHY motivate us to further analyze psychotherapy interactions by fusing acoustic and linguistic features in future works.

\section{Acknowledgements}
This research is partially supported by the Sustainable Research Fund of the Chinese University of Hong Kong (CUHK) and an ECS grant from the Hong Kong Research Grants Council (Ref.: 24604317).

\bibliographystyle{IEEEtran}

\bibliography{mybib}

% \begin{thebibliography}{9}
% \bibitem[1]{Davis80-COP}
%   S.\ B.\ Davis and P.\ Mermelstein,
%   ``Comparison of parametric representation for monosyllabic word recognition in continuously spoken sentences,''
%   \textit{IEEE Transactions on Acoustics, Speech and Signal Processing}, vol.~28, no.~4, pp.~357--366, 1980.
% \bibitem[2]{Rabiner89-ATO}
%   L.\ R.\ Rabiner,
%   ``A tutorial on hidden Markov models and selected applications in speech recognition,''
%   \textit{Proceedings of the IEEE}, vol.~77, no.~2, pp.~257-286, 1989.
% \bibitem[3]{Hastie09-TEO}
%   T.\ Hastie, R.\ Tibshirani, and J.\ Friedman,
%   \textit{The Elements of Statistical Learning -- Data Mining, Inference, and Prediction}.
%   New York: Springer, 2009.
% \bibitem[4]{YourName17-XXX}
%   F.\ Lastname1, F.\ Lastname2, and F.\ Lastname3,
%   ``Title of your INTERSPEECH 2022 publication,''
%   in \textit{Interspeech 2022 -- 23\textsuperscript{rd} Annual Conference of the International Speech Communication Association, September 18-22, Incheon, Korea, Proceedings, Proceedings}, 2022, pp.~100--104.
% \end{thebibliography}

\end{document}